\documentstyle[aps,prb,wasysym,graphicx]{revtex}

\begin{document}

\twocolumn[\hsize\textwidth\columnwidth\hsize\csname
@twocolumnfalse\endcsname

\draft{}

\title{Anisotropy of the Mobility of Pentacene from Frustration}

\author{Gilles A. de Wijs$^{1,2}$, Christine C. Mattheus$^1$,
 Robert A. de Groot$^{1,2}$, and Thomas T.M. Palstra$^1$}

\address{
$^1$Laboratory of Chemical Physics, Materials Science Centre,\\
Nijenborgh~4, NL-9747~AG Groningen, The~Netherlands\\
$^2$Computational Materials Science, FOM Institute for Condensed Matter,\\
Toernooiveld~1, NL-6525~ED Nijmegen, The~Netherlands\\
}

\date{\today}
\maketitle

\begin{abstract}
The bandstructure of pentacene is calculated using first-principles density
functional theory. A large anisotropy of the hole and electron effective
masses within the molecular planes is found. The band dispersion of the HOMO
and the LUMO is analyzed with the help of a tight-binding fit. The anisotropy
is shown to be intimately related to the herringbone structure.
\end{abstract}
\pacs{Keywords: pentacene, bandstructure, effective mass, anisotropy, mobility}

\vskip2pc]

\narrowtext

\section{Introduction}

Charge transport in organic molecular conductors has gained prominent
interest as it is expected that in ultra-clean materials band transport
with high electronic mobilities can be obtained. Recent progress in
synthetic procedures have minimized mobility reducing factors, such as
disorder, Frenkel defects, and Schottky defects. Thus, mobilities up to
$\sim$~1~cm$^2$/Vs have been obtained in thin film devices and in single
crystals at room temperature.\cite{nelson,lin,karl,buckyb}

Pentacene has gained prominent interest because of a number of favorable
materials properties. Large single crystals can be easily grown by vapor
transport,\cite{laudise} and the material can be handled in air. The
reactivity of pentacene to impurity levels of oxygen and hydrogen is
often underestimated, but can be minimized using appropriate synthetic
conditions.\cite{dihydro} Despite the presence of at least four
different polymorphs,\cite{mattheuspoly} there are no structural phase
transitions observed for the most stable polymorph, the single crystal
polymorph.\cite{holmes3,mattheus,siegrist2} The layered crystal
structure indicates a highly two-dimensional charge transport.

In this paper we use first-principles density functional theory (DFT) to
calculate the bandstructure of pentacene. We find a very large
anisotropy of the effective mass within the layers, resulting from a 
frustration of the electron transfer along the crystallographic {\bf b}-axis.
This frustration is rationalized using a tight-binding fit of the band
structure, and is intimately related to the herringbone geometry of the
crystal structure of pentacene. Our results imply that the mobility is
strongly directionally dependent within the layers.

\section{Structure}
\label{structure}

\begin{figure}
\includegraphics[width=80mm]{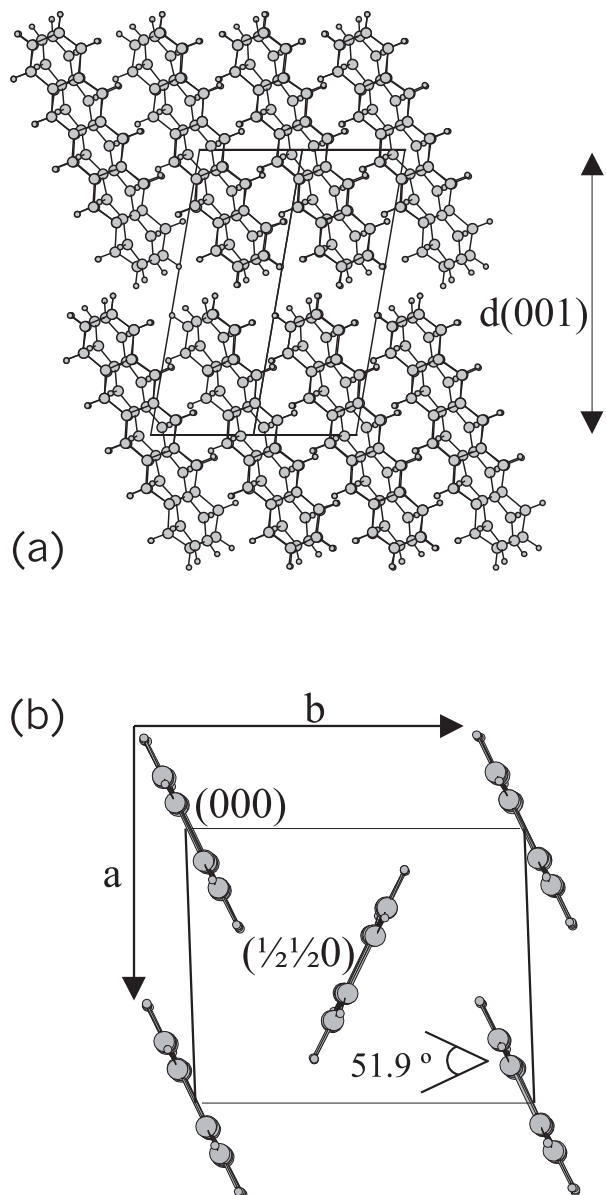}

\caption{{The crystal structure of pentacane.
  (a) Stacked layers of pentacene molecules,
  viewed along the $[1 \overline{1} 0]$-axis.
  A unit cell is also indicated.
  (b) Projection of the pentacene crystal structure
  and its unit cell vectors {\bf a} and {\bf b}
  on a plane perpendicular to the long molecular axes (LMA).
  The angle between the molecules is indicated.
  The herringbone arrangement is evident.}}
\label{strc}
\end{figure}

The crystal structure of bulk pentacene, depicted in Fig.~\ref{strc}, has been
reported by several groups.\cite{holmes3,mattheus,siegrist2}
It is a layered structure, triclinic, with spacegroup $P\bar{1}$.
The molecular layers consist of pentacene molecules
arranged in a herringbone pattern. We chose the unit vectors following
Ref.~\onlinecite{mattheus}, with {\bf a} and {\bf b} in the molecular plane
and {\bf c} inter-relating the molecular planes. The unit cell contains two
inequivalent molecules. The molecules are not perpendicular to the
molecular planes, but are at finite angles. This slant can be understood
as a relative shift of the molecules. Starting from one molecule, moving
along $({\bf a}+{\bf b})/2$ one arrives at an another (inequivalent) molecule,
that is shifted along its longest molecular axis (LMA) by 2.13~\AA,
i.e.\ approximately the length of one ring in the pentacene
molecule. Going along $({\bf a}-{\bf b})/2$ the shift is a mere 0.12~\AA, i.e.\ it
is negligible. This ``staircase'' pattern is indicated in Fig~\ref{strc}(b), which
shows a projection of a molecular layer along the LMA.

\section{Technical details}
\label{technical}

The bandstructure has been calculated using density functional theory
in the generalized gradient approximation (GGA).\cite{perdew}
The {\it ab initio} total-energy and molecular-dynamics program VASP
(Vienna {\it Ab initio} Simulation Program) was
used.\cite{vasp1,vasp2,vasp3,vasp4}
Electron-ion interactions were described using the projector augmented
wave method.\cite{bloechlpaw,kressepaw}
The kinetic energy cutoff on the wave function expansion was 500~eV.

The self-consistent calculations were carried out using the experimental
positions and cell (determination at 90~K from Ref.~\onlinecite{mattheus})
with a 4$\times$4$\times$2 {\bf k}-point mesh.\cite{monk} Convergence
was tested with a 6$\times$6$\times$4 mesh and found to be sufficient.

The bandstructure of the single molecular layer was calculated also from
a 3D crystal, but with an extra vacuum region of 6~\AA\ thickness inserted
between the layers. Thus interactions between the layers were avoided.
Moreover, the ${\bf c}$-axis was chosen perpendicular to the ${\bf ab}$
plane, i.e.\ the lattice was transformed to that of a monoclinic crystal.

\section{Results}
\label{results}

\begin{figure}
\includegraphics[width=80mm]{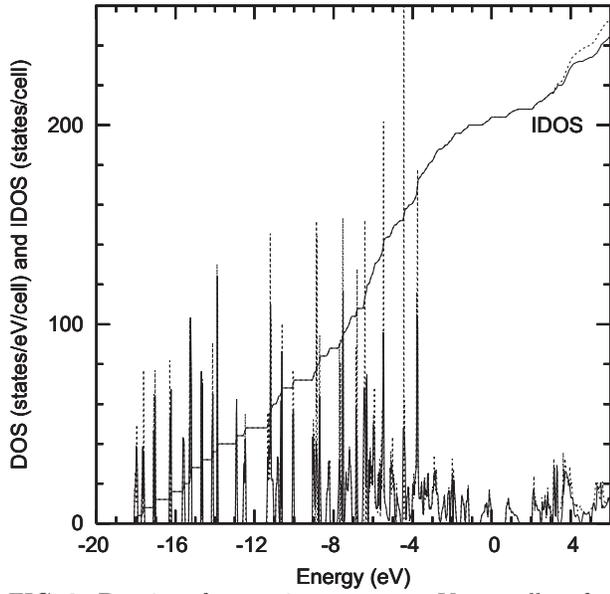}
   \caption{{Density of states in states per eV per cell
   as function of the energy. The energy is chosen zero
   at the top of the valence band. Solid line: single crystal,
   dotted line: cell with the molecules 6~\AA\ apart. Both
   curves are calculated using a 4$\times$4$\times$2 mesh.
   IDOS is the integrated DOS in states(electrons) per cell.}}
\label{dos}
\end{figure}

Fig.~\ref{dos} displays the calculated density of states (DOS) of pentacene.
All bands below the Fermi level are completely filled. The unoccupied
bands are separated from the valence bands by a band gap of 0.7~eV.
This is consistent with experiment, as undoped pentacene is an insulator.
The experimental band gap is 2.2~eV.\cite{silinsh} However, an underestimation
of the band gap is not unusual for GGA/DFT and, in general, doesn't much
affect the other features of the electronic structure.
The DOS peak of the HOMO (just below 0~eV) and of the LUMO (at 1~eV) are both
well separated from the other peaks. Each of them doesn't mix with bands
derived from other molecular states. The calculated HOMO and LUMO band
widths are $\sim$~600 and $\sim$~700~meV respectively.

Another, recent theoretical calculation gives values of 608 and 588~meV
respectively.\cite{cornil} This compares well with our results, although
we find a bit wider LUMO band.

Both the DOS of the single layer and of the complete 3D crystal are plotted.
On the scale of the plot, differences are only apparent above 3~eV. The
electronic structure of the single layer should be a good first approximation to
that of the complete crystal. Its bandstructure is only 2D, which greatly
facilitates the interpretation.

\subsection{Bandstructure of a Single Layer}
\label{single}

\begin{figure}
\includegraphics[width=80mm]{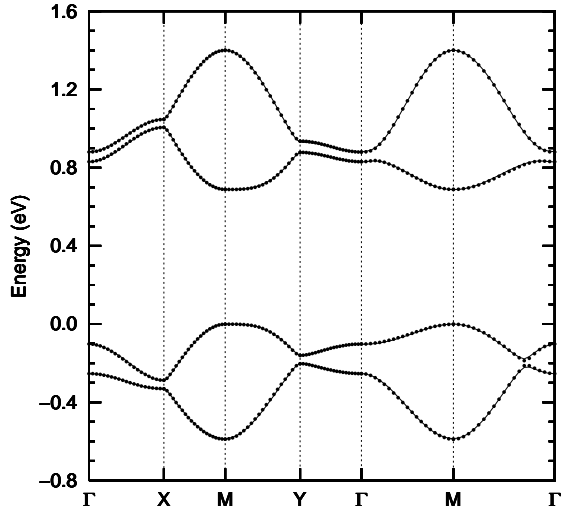}
   \caption{{The band structure of one layer of pentacene
   molecules, with a layer spacing of 6~\AA. The energy is
   chosen to be zero at the top of the valence band. The energies are plotted
   along lines in the first Brillouin zone, connecting the points $\Gamma =
(0,0)$,
   $X = (a^*/2, 0)$, $M = (a^*/2, b^*/2)$, and
   $Y = (0, b^*/2)$. From $\Gamma$ to $M$ is in the direction of
   $1/2{\bf (a^*+b^*)}$, from $M$ to $\Gamma$ is from $1/2{\bf (a^*-b^*)}$ to
   $\Gamma$. The solid line is the tight binding fit.}}
\label{band2D}
\end{figure}

The band structure near the Fermi level of a molecular layer is depicted in
Fig.~\ref{band2D}. Both the HOMO and the LUMO complex consist of two bands,
as there are two molecules in the unit cell. For both, the maximum and minimum
are at M [($\frac{1}{2}$,$\frac{1}{2}$)]. Therefore, upon doping (both
with electrons and with holes) the
charge carriers will be created in the vicinity of this point. Their
effective mass is strongly directionally dependent. In particular, note
the flatness of the bands near the extrema for the direction M-Y. These
observations will be further elaborated upon below.

\begin{figure}
\includegraphics[width=80mm]{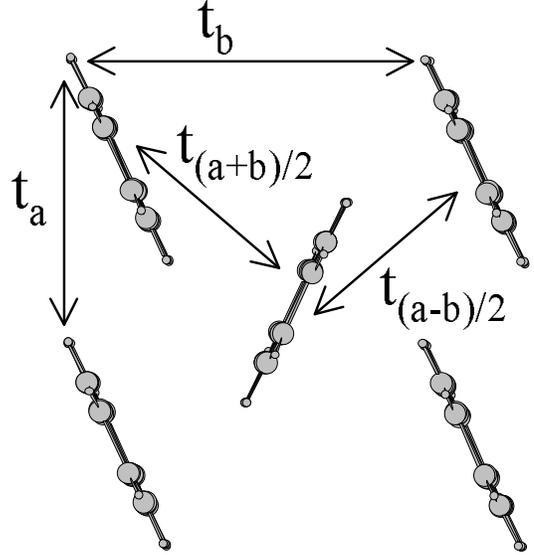}
   \caption{{The pentacene $ab$-plane, projected along the long molecular axis.
   The directions of the different overlap integrals are indicated.}}
\label{tdir}
\end{figure}

Since the band complexes are well separated from the other bands, we can
attempt a tight-binding (TB) fit, for both the HOMO and the LUMO separately.
The fit parameters are the transfer integrals $t$ between the molecular
orbitals.
Some of these are shown in Fig.~\ref{tdir}.
As we have two molecules, a $2 \times 2$ matrix will have to be diagonalized:
\begin{equation}
\left[ \begin{array}{cc}
      \varepsilon({\bf k}) - e - T_{1} & T_{2} \\ T_{2}   & \varepsilon

({\bf k}) - T_{1}
             \end{array} \right]
\label{tbdiag}
\end{equation}
Here $T_1$ and $T_2$ denote the sums of the diagonal and off-diagonal transfer
integrals, multiplied with the appropriate phase factors, respectively.
$\varepsilon({\bf k})$ are the eigenvalues and $e$ accounts for the fact the
the molecules are inequivalent. $\varepsilon({\bf k})$ is required to follow
the DFT result of Fig.~\ref{band2D} as close as possible.

As is evident from the figure, the fit follows the DFT result very well.
All transfer integrals are listed in Table~\ref{transfer}. By far the most important
ones are $t_{\bf a}$, $t_{({\bf a} + {\bf b})/2}$ and
$t_{({\bf a} - {\bf b})/2}$.
For the HOMO (LUMO) the values are: $t_{\bf a}$~=~31~(-41)~meV,
$t_{({\bf a} + {\bf b})/2}$~=~-56~(-90)~meV and
$t_{({\bf a} - {\bf b})/2}$~=~91~(90)~meV. All the others are very small.

\begin{figure}
\includegraphics[width=80mm]{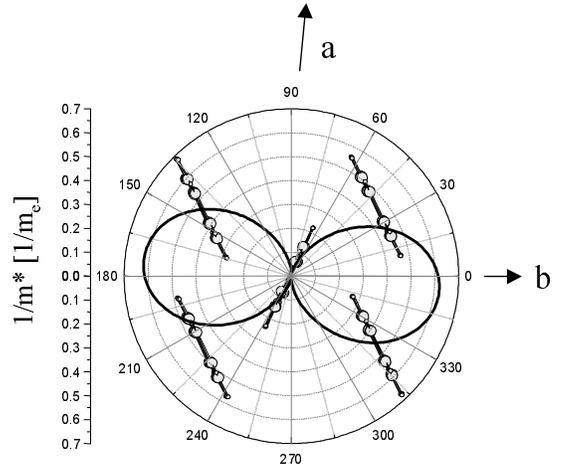}
   \caption{{The reciprocal effective mass as function of
   the crystallographic direction. Calculations are based on the
   tight-binding fit of the band structure, which is calculated
   for one layer of pentacene molecules. The pentacene molecules
   are shown as reference to the crystallographic directions.}}
\label{effmass}
\end{figure}

The effective mass is easily calculated with the help of the analytic
expression for the fit.
In Fig.~\ref{effmass} the hole effective mass $m^*$, calculated from:
\begin{equation}
 - \frac{\partial^{2}\varepsilon(k)}{\partial k^{2}} =\frac{\hbar^{2}}{m^{*}}
\label{efmsform}
\end{equation}
is plotted for all directions in the plane.
Note the large anisotropy. The largest value, 58~m$_{\rm e}$, occurs for the
${\bf a}$ direction.

\begin{figure}
\includegraphics[width=80mm]{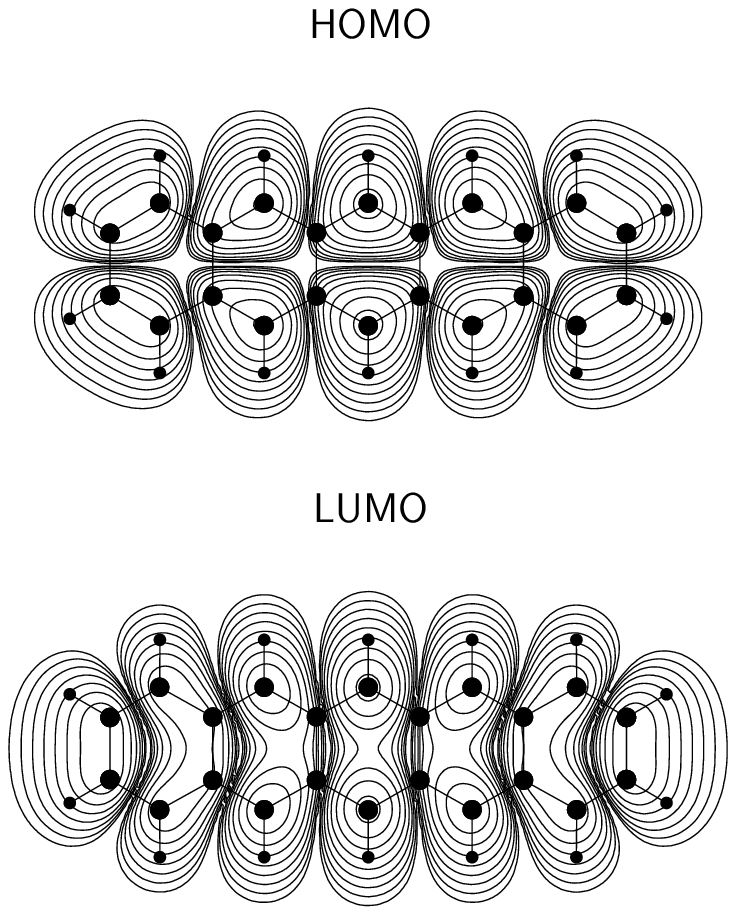}
\caption{{Calculated electron densities of the HOMO and the LUMO in a
  plane above the molecular plane.
  The large and small black circles indicate the position of the C-atoms and
  H-atoms, respectively. The lines are on a logarithmic scale. The
  lines lie at 0.0001~$\cdot$~10$^{n/3}$~elec/\AA$^{3}$, ~$n \geq 1$.
  The outermost lines have the same electron density.}}
\label{homolumo}
\end{figure}

Below we'll further discuss the band structure near the M point (i.e.\ the
point where the valence band maximum and conduction band minimum occur)
and explain
the mechanism behind the large anisotropy. The molecular orbitals,
HOMO and LUMO, are plotted in Fig.~\ref{homolumo}.\@ They are both
$\pi$-systems, having
a nodal plane coinciding with the plane of the molecule. Along the
LMA the HOMO alternatingly has maxima and minima that are slightly
more apart than the length of one ring-unit of the molecule. This pattern
leads to 4 nodal surfaces ``perpendicular'' to the LMA.
Additionally there's one more nodal plane containing the LMA.

The LUMO doesn't have such a nodal plane. It has two more nodal surfaces 
``perpendicular'' to
its LMA. Again the wave function extrema are approximately one
ring-unit apart.

\begin{figure}
\includegraphics[width=80mm]{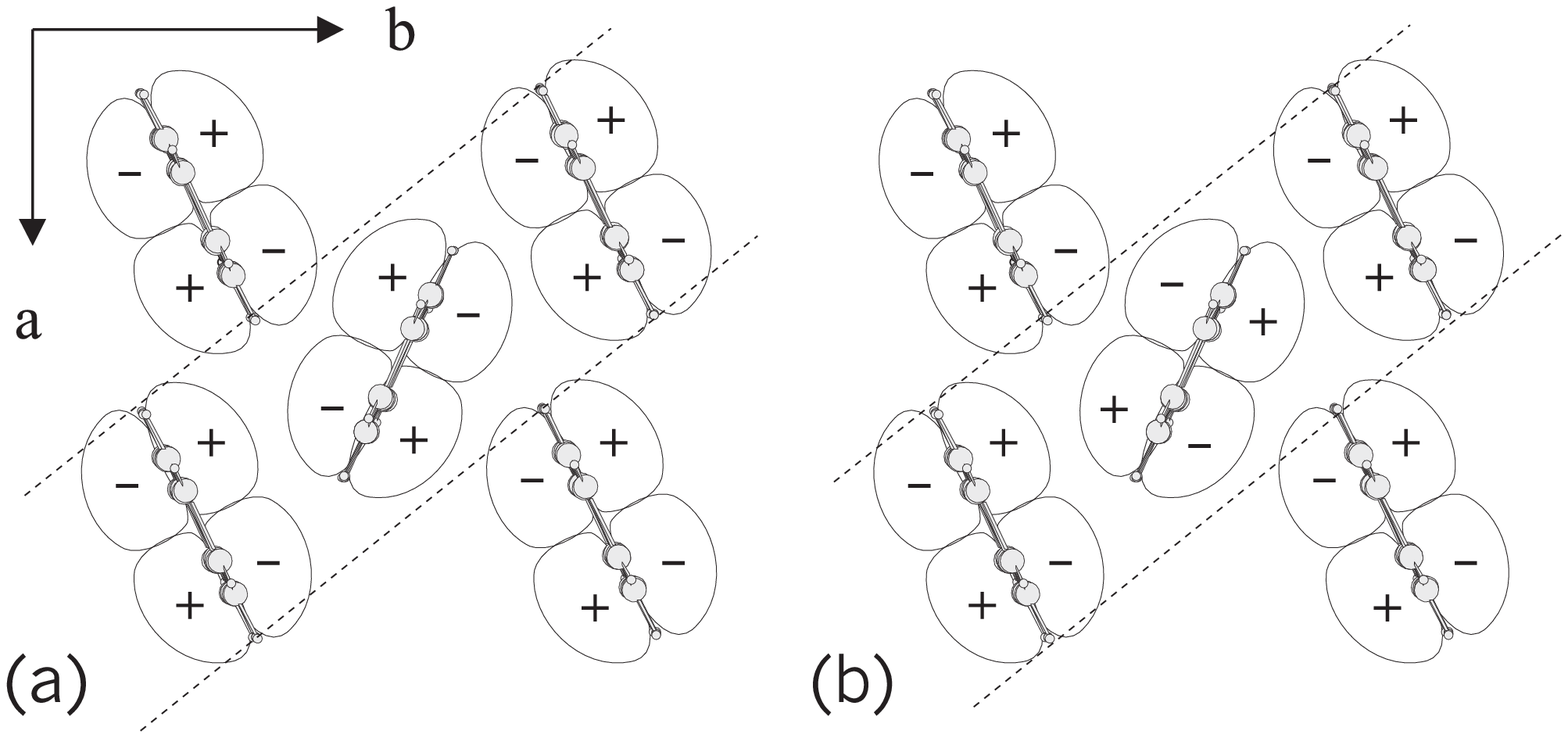}
 \caption{{Crystal orbitals of the HOMO complex 
   at M [($\frac{1}{2}$,$\frac{1}{2}$)], viewed along the LMA, in a plane
   perpendicular to the LMA.
   The HOMO wave functions
   are schematically depicted with the + and - signs to indicate the
   phase. (a) The most antibonding
   situation, as is the case at the top of the valence band. (b)
   All orbital overlaps are bonding, forming the bottom of the valence
   band. The dashed lines indicate the shifts between the molecules.}}
\label{schema}
\end{figure}

Fig.~\ref{schema} shows a schematic representation,
in a plane perpendicular to the LMA,
of the HOMO-derived crystal orbitals at the M point.
At M the phase of the wave function changes by 180$^\circ$,
both for a translation over ${\bf a}$ and ${\bf b}$.
However, in Fig.~\ref{schema} all 4 molecules at the corners have the
same phase. This is because the staircase (slant) in the layer gives
rise to a shift of the molecule at each step. This results (see the
preceding paragraph) in an additional phase change of approximately
180$^\circ$.

Depending on the phase of the central molecule, either the
low or high band at M is selected. In Fig.~\ref{schema}(b)
all three important
transfer integrals give a bonding contribution. Indeed, this is the
most bonding eigenstate within the band complex and thus corresponds
to the minimum at M. Fig.~\ref{schema}(a) depicts the most antibonding situation
(the top of the valence band): The
transfers $({\bf a} + {\bf b})/2$ and $({\bf a} - {\bf b})/2$ are maximally
anti-bonding. Transfer ${\bf a}$ is still bonding. Due to frustration no
solution with anti-bonding overlap for all three transfers is possible.

The frustration can also explain the large anisotropy in the effective mass.
Moving along ${\bf a}$ ($\approx$~${\bf a}^{*}$) two conduction channels are
possible: One directly {\it via} $t_{\bf a}$, the other {\it via} a ``double
hop'' [$t_{({\bf a} + {\bf b})/2}$ and $t_{({\bf a} - {\bf b})/2}$].
Moving away from M, the double hop will force the bands in a downward curvature,
which is counteracted by the direct channel. This explains why the band becomes
flat. By contrast, moving along ${\bf b}^{*}$ in reciprocal space, the
channel along ${\bf a}$ drops out and a
cosine-shaped band results, indicative of alternate $({\bf a} + {\bf b})/2$
and $({\bf a} - {\bf b})/2$ hops.

The LUMO bands mimic the HOMO bands in a very special way: They
appear as the ``mirror'' image of the HOMO bands on the energy axis
(Fig.~\ref{band2D}).
This peculiar feature is due to the LUMO's
slightly different nodal structure (it lacks the second mirror plane containing
the LMA). Therefore the top point at M has all three transfers antibonding,
whereas frustration now occurs for the most bonding point at the bottom of the
conduction band (where the ${\bf a}$ ``hop'' is still antibonding).
Thus the large anisotropy of the hole effective mass is ``mirrored'' in
the electron effective mass.
The analogy between the HOMO and LUMO band complex is not complete: E.g.,
the avoided crossing along M-$\Gamma$ only occurs for the HOMO and the total
width of the LUMO complex is slightly larger than that of the HOMO complex.

\subsection{Bandstructure of the 3D Crystal}
\label{3dcrystal}

\begin{figure}
\includegraphics[width=80mm]{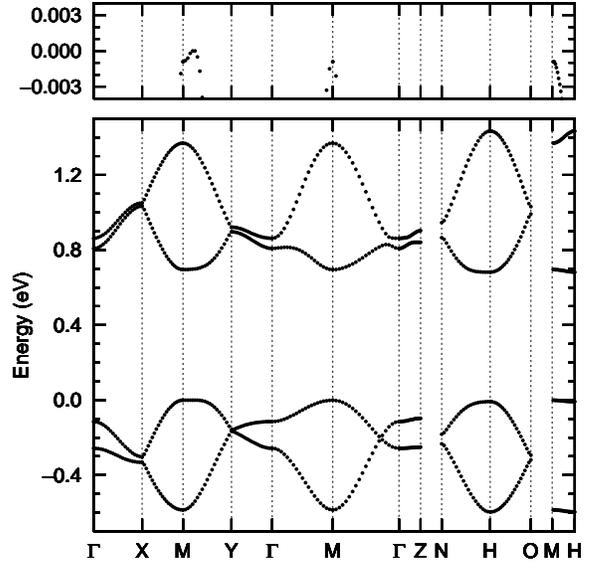}
   \caption{{The band structure of single crystalline pentacene
   The energy is chosen to zero at the top of the valence band. The
   energies are plotted along lines in the first Brillouin zone,
   connecting the points $\Gamma = (0,0,0)$, $X = (a^*/2,0,0)$,
   $M = (a^*/2,b^*/2,0)$, $Y = (0,b^*/2,0)$, $Z = (0,0,c^*/2)$,
   $N = (0,b^*/2,c^*/2)$, $H = (a^*/2,b^*/2,c^*/2)$ and $O = (a^*/2,0,c^*/2)$.
   From $\Gamma$ to $M$ is in the direction of $1/2{\bf (a^*+b^*)}$, from $M$
   to $\Gamma$ is from $1/2{\bf (a^*-b^*)}$ to $\Gamma$. The solid line is the
   tight binding fit. The top graph shows an enlargement of the top of
   the valence band.}}
\label{band3D}
\end{figure}

The band structure of the full 3D crystal (Fig.~\ref{band3D}) strongly
resembles that of the 2D layer. The shape of the bands for the most
dispersive, ${\bf a}^*$ and ${\bf b}^*$ directions is essentially the
same. Some differences are apparent though: (a) the top of the valence
band is not at M [($\frac{1}{2}$,$\frac{1}{2}$,0)], but at (0.375, 0.5, 0.075).
Moreover, as is evident from the top panel of Fig.~\ref{band3D}, the band
doesn't have a parabolic shape. The deviations are small however.

The conduction band minimum now occurs near H.
(b) A small band dispersion occurs along ${\bf c}^*$.

\begin{figure}
\includegraphics[width=80mm]{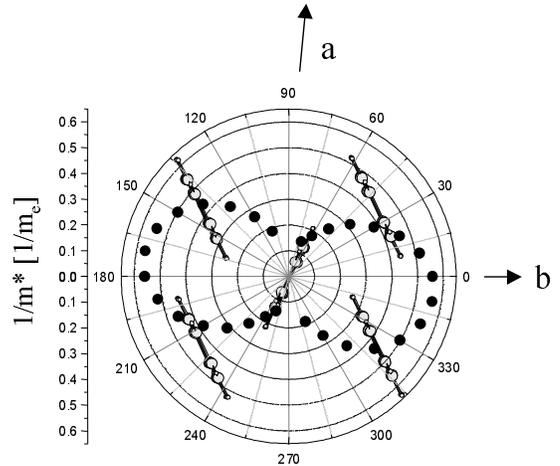}
\caption{{The reciprocal effective mass as function of
   the crystallographic direction. Calculations are based on the
   band structure of single crystalline pentacene. The pentacene
   molecules are shown as reference to the crystallographic directions.}}
\label{effmass3D}
\end{figure}

Fig.~\ref{effmass3D} depicts the hole effective mass, directly calculated
from the band structure using finite differences. We cannot rely on a TB
fit, as it cannot capture the bump at (0.375, 0.5, 0.075). For directions
close to ${\bf a}^*$ the effective mass was not calculated, due to the
locally non-parabolic
nature of the band. For the other directions the picture is very similar to
the 2D analogon. For ${\bf c}^*$ we obtain: $m^*_{{\bf c}^*}$~=~5.2~m$_e$.

Our results may be compared with a recent extended H{\"{u}}ckel (EH)
band structure
calculation on the pentacene single crystal.\cite{haddon}
The agreement between the EH and our first-principles results is quite good.
The EH study reports the band structure for a limited number of lines in the BZ
($\Gamma$-X, $\Gamma$-Y, $\Gamma$-Z) and therefore doesn't consider the
special situation near M.

\section{Discussion \&\ Conclusions}
\label{conclusions}

Band structure calculations for the pentacene single crystal reveal HOMO and
LUMO bandwidths of $\sim$~600 and 700~meV respectively. Consistently, (hole)
effective masses as low as 1.7~m$_e$ (for m$_{b^*}$) are found. Even
$m^*_{{\bf c}^*}$ (5.2~m$_e$) is rather low, in spite of the apparent
flatness of the bands for the ${\bf c}^*$ direction.
Within the molecular layers a competition between different conduction
paths results in a large anisotropy of both the hole and the
electron effective mass, the charge carriers being orders of magnitude
heavier for the ${\bf a}^*$ (than for the ${\bf b}^*$ direction).

Our calculations do not take into account the coupling with lattice
degrees of freedom. At lower temperatures the electron-phonon coupling
may renormalize the charge-carrier's bandwidth. At higher temperatures
the bandwidth renormalization and mobility reduction may differ for
different directions, thus affecting the anisotropy. For anthracene,
recent first-principles calculations indeed show a directionally
dependent bandwidth renormalization.\cite{bobbert2}
However, as it is so pronounced, we expect much of the effective mass
anisotropy to survive, also at higher temperatures, as long as coherent
band-like transport is dominant.

\section*{Acknowledgments}

We thank Dr.\ P.A.\ Bobbert for useful discussions and making available
Ref.~\onlinecite{bobbert2} prior to publication.
This work is part of the research programme of the
'Stichting voor Fundamenteel Onderzoek der Materie (FOM)',
which is financially supported by the
'Nederlandse Organisatie voor Wetenschappelijk Onderzoek (NWO)'.

\begin{table}
\caption{The following table lists the parameters for the single
layer TB fits.}
\begin{tabular}{ldd}
                & HOMO    & LUMO   \\
\hline
 $e$ (eV)       &  0.042  &  0.047  \\
 $t_{a}$        &  0.061  & -0.041  \\
 $t_{b}$        & -0.004  & -0.008  \\
 $t_{2a}$       & -0.001  &  0.000  \\
 $t_{a+b}$      &  0.003  & -0.002  \\
 $t_{a-b}$      &  0.000  & -0.002  \\
 $t_{(a+b)/2}$  & -0.056  & -0.090  \\
 $t_{(a-b)/2}$  &  0.091  &  0.090  \\
 $t_{3(a+b)/2}$ & -0.001  &  0.000  \\
 $t_{3(a-b)/2}$ &  0.001  &  0.000  \\
 $t_{(3a+b)/2}$ &  0.000  & -0.001  \\
 $t_{(3a-b)/2}$ &  0.001  &  0.001  \\
 $t_{ae}$       &  0.000  &  0.003  \\
\end{tabular}
\label{transfer}
\end{table}

\end{document}